\documentclass[12pt,epsf]{article}
\usepackage{amssymb,amsmath}
\usepackage{graphicx}
\usepackage[dvips]{color}

\newcommand{\be}{\begin{equation}}
\newcommand{\ee}{\end{equation}}
\newcommand{\bea}{\begin{eqnarray}}
\newcommand{\eea}{\end{eqnarray}}
\newcommand{\beas}{\begin{eqnarray*}}
\newcommand{\eeas}{\end{eqnarray*}}
\newcommand{\ba}{\begin{array}}
\newcommand{\ea}{\end{array}}

\newcommand{\nbox}{{\,\lower0.9pt\vbox{\hrule \hbox{\vrule height 0.2 cm \hskip 0.19 cm \vrule height 0.2 cm}\hrule}\,}}

\def\href#1#2{#2}
\input{tcilatex}
\begin{document}

\begin{titlepage}
\hfill
\vbox{
    \halign{#\hfil         \cr
           } % end of \halign
      }  % end of \vbox

\hbox to \hsize{{}\hss \vtop{ \hbox{}

}}

\begin{flushright}

\end{flushright}

\vspace*{20mm}
\begin{center}

{\Large \textbf{Giant gravitons and correlators} }

{\Large \vspace{10mm} }

\vspace*{15mm} \vspace*{1mm} Hai Lin{\footnote{e-mail:
hai.lin@usc.es}}

{\normalsize \vspace{7mm} }

{\normalsize \vspace{0.2cm} }

{\normalsize \emph{\textit{Department of Particle Physics,
Facultad de
Fisica, \\
Universidad de Santiago de Compostela, 15782 Santiago de
Compostela, Spain
\\
}} }

\end{center}

\vspace{10mm}

\begin{abstract}

\vspace{3mm}

We calculate non-extremal correlators of Schur polynomials and
single trace operators. We analyse their dual descriptions from
the approach of the variation of DBI and WZ actions of the giant
gravitons. We show a regularization procedure under which the
extremal correlators of Schur polynomials and single trace
operators match exactly with string theory computation. Other
aspects of the extremal and non-extremal correlators are also
discussed.

\end{abstract}

\end{titlepage}

\vskip 1cm

\section{Introduction}

\vspace{1pt}\renewcommand{\theequation}{1.\arabic{equation}} %
\setcounter{equation}{0}\bigskip

In this paper we study correlators for giant gravitons in AdS/CFT \cite%
{Maldacena:1997re},\cite{Gubser:1998bc},\cite{Witten:1998qj}. The giant
graviton in AdS space is a brane that grows large in AdS or in sphere
directions \cite{McGreevy:2000cw},\cite{Grisaru:2000zn},\cite%
{Hashimoto:2000zp}. They can be described by operators constructed by Schur
polynomials and labelled by Young diagrams \cite{Corley:2001zk}, see also
related discussion \cite{Balasubramanian:2001nh},\cite{Berenstein:2004kk}.
The giant graviton which grows in sphere directions are described by
operators in representation $1^{k}$ and those grows in AdS space are
described by operators in representation $k$.

The brane in AdS space couples with background gravitational field via the
DBI and WZ action. It is nice to consider correlators between two heavy
operators, such as those correspond to giant gravitons, and other light
operator corresponding to supergravity field. Recently, \cite{Bissi:2011dc}
and \cite{Caputa:2012yj}\ studied these correlators and compared them with
the variation of the DBI and WZ action describing their interaction. In
particular, \cite{Bissi:2011dc} analysed variation of the action, and found
that there is a disagreement in matching the extremal correlators. In
particular, \cite{Caputa:2012yj} analysed non-extremal correlators and found
that a large class of them agree with the gauge theory computation.

In this paper we will discuss the problem of matching the extremal
correlators and show a regularization procedure under which the string
theory computation and gauge theory computation agrees exactly. We also
study other class of non-extremal correlators in both the gauge theory and
the string theory perspectives and found agreement.

The organization of this paper is as follows. In section 2, we study the
correlator of two Schur polynomials and one single trace operator, in
particular the non-extremal correlators. In section 3.1, we perform detailed
analysis of the DBI and WZ variations of \cite{Bissi:2011dc}, and show a
regularization procedure that correctly gives the answer that exactly equals
the answer from the gauge theory computation. In section 3.2, we analyze
non-extremal correlators and show agreements nontrivially with the gauge
theory computation. In particular, the $k/N$ dependence are nicely captured
in the string theory computation. In section 3.3, we analyse AdS giant.
Finally, in section 4, we briefly discuss the results and conclusions.

%%%%%%%%%%%%%%%%%%%%%%%%%%%%%%%%%%%%%%%%%%%%%%%%%%%%%%%%%%%%%%%%%%%%
%%%%%%%%%%%%%%%%%%%%%%%%%%%%%%%%%%%%%%%%%%%%%%%%%%%%%%%%%%%%%%%%%%%%
%%%%%%%%%%%%%%%%%%%%%%%%%%%%%%%%%%%%%%%%%%%%%%%%%%%%%%%%%%%%%%%%%%%%
%%%%%%%%%%%%%%%%%%%%%%%%%%%%%%%%%%%%%%%%%%%%%%%%%%%%%%%%%%%%%%%%%%%%

\section{\protect\bigskip Correlators with giant gravitons in gauge theory}

\renewcommand{\theequation}{2.\arabic{equation}} \setcounter{equation}{0}%
\vspace{1pt}

\subsection{Schur polynomials and correlators}

\bigskip \label{sec_correlator_01}

The Schur Polynomial which corresponds to the giant graviton is defined as
\cite{Corley:2001zk}%
\begin{equation}
\chi _{R}(Z)={\frac{1}{k!}}\sum_{\sigma \in S_{k}}\chi _{R}(\sigma
)Z_{i_{\sigma (1)}}^{i_{1}}\cdots Z_{i_{\sigma (k-1)}}^{i_{k-1}}Z_{i_{\sigma
(k)}}^{i_{k}},
\end{equation}%
\begin{equation}
\langle \chi _{R}(Z)\chi _{T}(Z)^{\dagger }\rangle =\frac{k!~\text{Dim(}R%
\text{)}}{d_{R}}\delta _{R,T}=\prod\limits_{i,j\in R}(N-i+j).
\end{equation}%
The Young diagram $R$ labelling the Schur Polynomial that we will consider
has $k$ boxes.

We will compute the correlator of two Schur Polynomials $\chi _{R},\chi
_{R^{\prime }}~$and another single trace operator of the form
\begin{equation}
{\small :}\mathrm{Tr}(Z^{j_{1}}Z^{\dagger }Z^{j_{2}}Z^{\dagger }\cdots
Z^{j_{l}}Z^{\dagger }){\small :}  \label{operator_Z+_ge}
\end{equation}%
where the total number of $Z$'s is $n=\sum_{i=1}^{l}j_{i},$ and there are
total number of $l$ $Z^{\dagger }$'s. The total $u(1)$ R-charge of the
operator is $j=n-l$. We will distinguish the case of non-adjacent $%
Z^{\dagger }$'s, and adjacent $Z^{\dagger }$'s, in the computation of
correlators of them with Schur polynomials. If $j_{i}$ is not zero, then the
two $Z^{\dagger }$'s from the two sides is not adjacent. Otherwise if there
is a number $j_{i}$ that is zero, the two $Z^{\dagger }$'s are adjacent, in
other words, \ we have a $(Z^{\dagger })^{2}.$ Similarly we have situations
where there is $(Z^{\dagger })^{3},(Z^{\dagger })^{4},(Z^{\dagger })^{5}$
and so on.

Now let's consider the correlator between two Schur polynomials labelled by $%
R\vdash k$ and $R^{+}\vdash k+n-l$ , where $n=j+l$, and the third operator
as in (\ref{operator_Z+_ge}). The limit we focus on here is that $k\sim
O(N),~$and $j,l\ll k,N.$ In other words, the giant graviton operator has
much larger R-charge than the other light operator.

We first consider the situation that there is no adjacent $Z^{\dagger }$'s
in the trace of (\ref{operator_Z+_ge}). Those type will be the main terms
when $l\ll j.~$Consider the Schwinger-Dyson equation
\begin{equation}
0=\int \left[ dZdZ^{\dagger }\right] {\frac{d}{dZ_{ij}}}\left(
(Z^{j+1}Z^{\dagger }Z)_{ij}\chi _{R}(Z)\chi _{R^{+}}(Z)^{\dagger }e^{-%
\mathrm{Tr}(ZZ^{\dagger })}\right)
\end{equation}%
which implies that (self contractions are subtracted out)%
\begin{equation}
\;\langle \chi _{R}(Z)\chi _{R^{+}}(Z)^{\dagger }{\small :}\mathrm{Tr}%
(Z^{j}(ZZ^{\dagger })^{2}{\small :}\rangle =\langle \left(
(Z^{j+1}Z^{\dagger }Z)_{ij}{\frac{d}{dZ_{ij}}}\chi _{R}(Z)\right) \chi
_{R^{+}}(Z)^{\dagger }\rangle .
\end{equation}%
Now consider another Schwinger-Dyson equation%
\begin{equation}
0=\int \left[ dZdZ^{\dagger }\right] {\frac{d}{dZ_{ij}}}\left(
(Z^{j+1})_{n_{1}}^{i}Z_{j}^{k_{2}}{\frac{d}{dZ_{n_{1}}^{k_{2}}}}\chi
_{R}(Z)\chi _{R^{+}}(Z)^{\dagger }e^{-\mathrm{Tr}(ZZ^{\dagger })}\right)
\end{equation}%
which then implies that (self contractions are subtracted out)
\begin{equation}
\langle \left( (Z^{j+1}Z^{\dagger }Z)_{ij}{\frac{d}{dZ_{ij}}}\chi
_{R}(Z)\right) \chi _{R^{+}}(Z)^{\dagger }\rangle \;=\langle \left(
(Z^{j}Z)_{n_{1}}^{k_{1}}{\frac{d}{dZ_{n_{2}}^{k_{1}}}}Z_{n_{2}}^{k_{2}}{%
\frac{d}{dZ_{n_{1}}^{k_{2}}}}\chi _{R}(Z)\right) \chi _{R^{+}}(Z)^{\dagger
}\rangle .
\end{equation}%
By using the similar method several times, we have that
\begin{equation}
\langle \chi _{R}(Z)\chi _{R^{+}}(Z)^{\dagger }{\small :}\mathrm{Tr}%
(Z^{j}(ZZ^{\dagger })^{l}{\small :}\rangle =\langle \left(
(Z^{j}Z)_{n_{1}}^{k_{1}}{\frac{d}{dZ_{n_{2}}^{k_{1}}}}Z_{n_{2}}^{k_{2}}{%
\frac{d}{dZ_{n_{3}}^{k_{2}}}}\cdots Z_{n_{l}}^{k_{l}}{\frac{d}{%
dZ_{n_{1}}^{k_{l}}}}\chi _{R}(Z)\right) \chi _{R^{+}}(Z)^{\dagger }\rangle .
\end{equation}

So we will compute
\begin{eqnarray}
&&(Z^{j}Z)_{n_{1}}^{k_{1}}{\frac{d}{dZ_{n_{2}}^{k_{1}}}}Z_{n_{2}}^{k_{2}}{%
\frac{d}{dZ_{n_{3}}^{k_{2}}}}\cdots Z_{n_{l}}^{k_{l}}{\frac{d}{%
dZ_{n_{1}}^{k_{l}}}}\chi _{R}(Z)  \notag \\
&=&{\frac{1}{(k-l)!}}\sum_{\sigma \in S_{k}}\chi _{R}(\sigma )Z_{i_{\sigma
(1)}}^{i_{1}}\cdots Z_{i_{\sigma (k-1)}}^{i_{k-1}}(Z^{j+1})_{i_{\sigma
(k)}}^{i_{k}}  \notag \\
&=&{\frac{1}{(k-l)!}}\sum_{\sigma \in S_{k}}\sum_{T\vdash \,k+j}\chi
_{R}(\sigma )\chi _{T}(\sigma (k,k+1,{\small \cdots },k+j)\,)\chi _{T}(Z) \\
&=&{\frac{k!}{(k-l)!d_{R}}}\sum_{T\vdash \,k+j}\mathrm{Tr}(P_{T\rightarrow
R}(k,k+1,{\small \cdots },k+j)\,)\chi _{T}(Z),  \label{d_Z_Chi_non-a}
\end{eqnarray}%
where $P_{T\rightarrow R}$ is a projector that projects to $R$ after
restricting to the $S_{k}$ subgroup and is zero if $T$ does not subduce $R$
after restricting to the $S_{k}$ subgroup. The trace is taken over the
carrier space of $T$. We have also used the fundamental orthogonality
relation to sum over $\sigma $. We therefore have
\begin{eqnarray}
\langle \chi _{R}(Z)\chi _{R^{+}}(Z)^{\dagger } {\small :}\mathrm{Tr}%
(Z^{j}(ZZ^{\dagger })^{l}{\small :}\rangle =\frac{k!}{(k-l)!d_{R}}\mathrm{Tr}%
(P_{R^{+}\rightarrow R}(k,k+1,\cdots ,k+j)\,)\langle \chi _{R^{+}}(Z)\chi
_{R^{+}}(Z)^{\dagger }\rangle .  \notag  \label{correlator_schur_non_a} \\
\end{eqnarray}%
The above expression is illustrated for the situation when $j_{1}=j+1,~$and $%
j_{i}=1$ for $i=2,...,l$.~Other situations without adjacent $Z^{\dagger }$'s
is similar, but with a different projector, and the the final answer is
similar to (\ref{correlator_schur_non_a}).\ \ In the regime $l\ll k,$ we
have that$\bigskip ~\frac{k!}{(k-l)!}\simeq k^{l}.~$This formula is the same
for both representation of $(k)$ and $(1^{k})$.

Now we look at the situation that there are adjacent $Z^{\dagger
}$'s$.$ For example, for the situation with $l$ adjacent
$Z^{\dagger }$'s, using the above method, we have that
\begin{equation}
\langle \chi _{R}(Z)\chi _{R^{+}}(Z)^{\dagger }{\small :}\mathrm{Tr}%
(Z^{j+l}Z^{\dagger }{}^{l}){\small :}\rangle =\langle \left(
(Z^{j+l})_{i}^{j}{\frac{d}{dZ_{n_{1}}^{j}}\frac{d}{dZ_{n_{2}}^{n_{1}}}}...{%
\frac{d}{dZ_{n_{l-1}}^{n_{l-2}}}\frac{d}{dZ_{i}^{n_{l-1}}}}\chi
_{R}(Z)\right) \chi _{R^{+}}(Z)^{\dagger }\rangle ,
\end{equation}%
and we evaluate
\begin{eqnarray}
&&(Z^{j+l})_{i}^{j}{\frac{d}{dZ_{n_{1}}^{j}}\frac{d}{dZ_{n_{2}}^{n_{1}}}}...{%
\frac{d}{dZ_{n_{l-1}}^{n_{l-2}}}\frac{d}{dZ_{i}^{n_{l-1}}}}\chi _{R}(Z)
\notag \\
&=&{\frac{1}{(k-l)!}}\sum_{\sigma \in S_{k}}\chi _{R}(\sigma )Z_{i_{\sigma
(1)}}^{i_{1}}\cdots Z_{i_{\sigma (k-1)}}^{i_{k-1}}(Z^{j+l})_{i_{\sigma
(k)}}^{i_{k-l+1}}\delta _{i_{\sigma (k-1)}}^{i_{k}}\delta _{i_{\sigma
(k-2)}}^{i_{k-1}}...\delta _{i_{\sigma (k-l+1)}}^{i_{k-l+2}}  \notag \\
&=&\prod_{\alpha =1}^{l-1}\oplus _{R_{\alpha }}c_{R_{\alpha -1}R_{\alpha }}{%
\frac{1}{(k-l)!}}\sum_{T\vdash k-j}\sum_{\sigma \in S_{k-l+1}}\chi
_{R^{\prime }}({\normalsize \sigma (k,k-1,k-2,...,k-l+1)})  \notag \\
&&\chi _{T}(\text{{\normalsize $\sigma $$(k-l+n,k-l+n+1,...,k-l+1)$}})\chi
_{T}(Z) \\
&=&\prod_{\alpha =1}^{l-1}\oplus _{R_{\alpha }}c_{R_{\alpha -1}R_{\alpha }}{%
\frac{(k-l+1)!}{(k-l)!d_{R^{\prime }}^{2}}}\sum_{T\vdash k-j}\mathrm{Tr}%
(P_{R\rightarrow R^{\prime }}{\normalsize (k,k-1,\cdots ,k-l+1)}{\scriptsize %
\,})  \notag \\
&&\mathrm{Tr}(P_{T\rightarrow R^{\prime }}{\normalsize %
(k-l+n,k-l+n+1,...,k-l+1)}{\scriptsize \,})\chi _{T}(Z).  \label{young__d^l_}
\end{eqnarray}%
In the above, $c_{R_{\alpha -1}R_{\alpha }}$ is the weight of the box that
is removed from Young diagram $R_{\alpha -1}$ to obtain Young diagram $%
R_{\alpha }.$ We denote $R_{0}=R,$ and $R_{l-1}=$ $R^{+}.$ In
other words, we need to remove $l-1$ boxes, if there is $l$
adjacent $Z^{\dagger }. $ Similarly, if there is a $(Z^{\dagger
})^{m}$, in other words, $m$ adjacent $Z^{\dagger }$'s, then one
needs to remove $m-1$ boxes from the
Young diagram $R.~$This situation is similar to \cite{Koch:2008cm}. From (%
\ref{young__d^l_}) we have that
\begin{eqnarray}
\langle \chi _{R}(Z)\chi _{R^{+}}(Z)^{\dagger } {\small :}\mathrm{Tr}%
(Z^{j+l}Z^{\dagger }{}^{l}){\small :}\rangle =\prod_{\alpha =1}^{l-1}\oplus
_{R_{\alpha }}c_{{\small {R_{\alpha -1}R_{\alpha }}}}(k-l+1)\frac{1}{%
d_{R^{\prime }}^{2}}\mathrm{Tr}(P_{R\rightarrow R^{\prime
}}(k,k-1,...,k-l+1)\,)  \notag  \label{d^l_R} \\
\mathrm{Tr}(P_{T\rightarrow R^{\prime
}}(k-l+n,k-l+n+1,...,k-l+1)\,)\langle \chi _{R^{+}}(Z)\chi
_{R^{+}}(Z)^{\dagger }\rangle .  \notag \\
\end{eqnarray}

The expression (\ref{young__d^l_}) is general for different representations
of $(k)$ or $(1^{k})$. For different representations, the difference is
mainly the weights $c_{R_{\alpha -1}R_{\alpha }}$. \ For sphere giant,
\begin{equation}
\prod_{\alpha =1}^{l-1}\oplus _{R_{\alpha }}c_{R_{\alpha -1}R_{\alpha
}}=\prod_{\alpha =1}^{l-1}(N-k+\alpha ),
\end{equation}%
and similarly for AdS giant,%
\begin{equation}
\prod_{\alpha =1}^{l-1}\oplus _{R_{\alpha }}c_{R_{\alpha -1}R_{\alpha
}}=\prod_{\alpha =1}^{l-1}(N+k-\alpha ).
\end{equation}%
In the regime $l\ll k,N$, we have that $\prod_{\alpha =1}^{l-1}(N+k-\alpha
)(k-l+1)\simeq k(N+k)^{l-1}.$

For the symmetric representation $R=(k)$ and $R^{+}=(k+j)=(k+n-l)$. These
irreducible representations are one-dimensional, and the projector is
trivial, and group elements are represented by $1$. So we find
\begin{eqnarray}
&&{\frac{\langle {\small :}\mathrm{Tr}(Z^{j}(ZZ^{\dagger })^{l}){\small :}%
\chi _{(k)}(Z)\chi _{(k+j)}(Z)^{\dagger }\rangle }{\sqrt{\langle \chi
_{(k+j)}(Z)\chi _{(k+j)}(Z)^{\dagger }\rangle \langle \chi _{(k)}(Z)\chi
_{(k)}(Z)^{\dagger }\rangle }\left\Vert {\small :}\mathrm{Tr}%
(Z^{j}(ZZ^{\dagger })^{l}){\small :}\right\Vert }}  \notag \\
&=&\left( {\frac{k}{N}}\right) ^{l}\left( 1+{\frac{k}{N}}\right) ^{\frac{j}{2%
}}\left( \frac{l+1}{j+2l+1}\right) ^{\frac{1}{2}},
\label{correlator_rep_k_non_a}
\end{eqnarray}%
where we reinstated the appropriate normalizations.

For the antisymmetric representation, $R=(1^{k})$ and $%
R^{+}=(1^{k+j})=(1^{k+n-l})$. These irreducible representations are one
dimensional, and the projector is trivial. There are sign factors due to
parity of permutations.\ The group elements in this representation are $%
\Gamma (\sigma )=(-1)^{\epsilon (\sigma )}$ where $\epsilon (\sigma )$ is
the parity of the permutation, and for the above permutations in (\ref%
{d_Z_Chi_non-a}), it gives factors $(-1)^{j}=(-1)^{n-l}$. Therefore we find
\begin{eqnarray}
&&{\frac{\langle {\small :}\mathrm{Tr}(Z^{j}(ZZ^{\dagger })^{l}){\small :}%
\chi _{(1^{k})}(Z)\chi _{(1^{k+j})}(Z)^{\dagger }\rangle }{\sqrt{\langle
\chi _{(1^{k+j})}(Z)\chi _{(1^{k+j})}(Z)^{\dagger }\rangle \langle \chi
_{(1^{k})}(Z)\chi _{(1^{k})}(Z)^{\dagger }\rangle }\left\Vert {\small :}%
\mathrm{Tr}(Z^{j}(ZZ^{\dagger })^{l}){\small :}\right\Vert }}  \notag \\
&=&\left( {\frac{k}{N}}\right) ^{l}\left( 1-{\frac{k}{N}}\right) ^{\frac{j}{2%
}}\left( \frac{l+1}{j+2l+1}\right) ^{\frac{1}{2}}.
\label{correlator_rep_1k_non_a}
\end{eqnarray}%
The factor $(-1)^{n-l}$ depends on conventions and can be removed. In most
of the discussions follows, we will remove this factor, but it can be
reinstated easily. \noindent

Similarly for the example of adjacent $Z^{\dagger }$'s, the answer for
symmetric and antisymmetric representations are respectively,
\begin{eqnarray}
&&{\frac{\langle {\small :}\mathrm{Tr}(Z^{j+l}Z^{\dagger }{}^{l}){\small :}%
\chi _{(k)}(Z)\chi _{(k+j)}(Z)^{\dagger }\rangle }{\sqrt{\langle \chi
_{(k+j)}(Z)\chi _{(k+j)}(Z)^{\dagger }\rangle \langle \chi _{(k)}(Z)\chi
_{(k)}(Z)^{\dagger }\rangle }\left\Vert {\small :}\mathrm{Tr}%
(Z^{j}(ZZ^{\dagger })^{l}){\small :}\right\Vert }}  \notag \\
&=&{\frac{k}{N}}\left( 1+{\frac{k}{N}}\right) ^{\frac{j}{2}+l-1}\left( \frac{%
l+1}{j+2l+1}\right) ^{\frac{1}{2}},
\end{eqnarray}%
\begin{eqnarray}
&&{\frac{\langle :\mathrm{Tr}(Z^{j+l}Z^{\dagger }{}^{l}):\chi
_{(1^{k})}(Z)\chi _{(1^{k+j})}(Z)^{\dagger }\rangle }{\sqrt{\langle \chi
_{(1^{k+j})}(Z)\chi _{(1^{k+j})}(Z)^{\dagger }\rangle \langle \chi
_{(1^{k})}(Z)\chi _{(1^{k})}(Z)^{\dagger }\rangle }\left\Vert {\small :}%
\mathrm{Tr}(Z^{j}(ZZ^{\dagger })^{l}){\small :}\right\Vert }}  \notag \\
&=&{\frac{k}{N}}\left( 1-{\frac{k}{N}}\right) ^{\frac{j}{2}+l-1}\left( \frac{%
l+1}{j+2l+1}\right) ^{\frac{1}{2}}.
\end{eqnarray}%
where in above notation $l\geqslant 1.$

Now we consider the situation that there are $l_{2}~$adjacent $Z^{\dagger }$%
's, each of which does not have a $Z$ on the left side; and there are $l_{1}~
$rest of the $Z^{\dagger }$'s, each of which has a $Z$ on the left side. In
other words, there are $l_{1}$ zero $j_{i}$'s$,$ and $l_{2}$ non-zero $j_{i}$%
's in (\ref{operator_Z+_ge}). The factor we get in this situation is ${k}%
^{l_{1}}\left( {N+k}\right) ^{l_{2}}$. This is because that for each $%
Z^{\dagger }~$that has a $Z$ on the left side$,$ they will give a factor ${k}
$; while for each $Z^{\dagger }~$that does not have a $Z$ on the left side,
they will give a factor $N+k$ instead. For example, ($Z^{\dagger })^{m}$%
~will give a factor ${k(N+k)}^{m-1}${, because there are }${m-1}${\ }$%
Z^{\dagger }$'s that does not have a $Z$ on the left side and{\ they involve
removing }${m-1~}${boxes, while the leftmost }$Z^{\dagger }$ gives the
factor $k$.{\ We can denote such operators as }$O_{j,l_{1},l_{2}},$ for
example,
\begin{equation}
O_{j,l_{1},l_{2}}={\small :}\mathrm{Tr}(Z^{j+l_{2}}(ZZ^{\dagger
}{})^{l_{1}}Z^{\dagger }{}^{l_{2}}){\small :}
\end{equation}%
for $l_{1}\geqslant 1$.$~$Therefore
\begin{eqnarray}
&&{\frac{\langle O_{j,l_{1},l_{2}}\chi _{(k)}(Z)\chi _{(k+j)}(Z)^{\dagger
}\rangle }{\sqrt{\langle \chi _{(k+j)}(Z)\chi _{(k+j)}(Z)^{\dagger }\rangle
\langle \chi _{(k)}(Z)\chi _{(k)}(Z)^{\dagger }\rangle }\left\Vert
O_{j,l_{1},l_{2}}\right\Vert }}  \notag \\
&=&\left( {\frac{k}{N}}\right) ^{l_{1}}\left( 1+{\frac{k}{N}}\right) ^{\frac{%
j}{2}+l_{2}}\left( \frac{l+1}{j+2l+1}\right) ^{\frac{1}{2}},
\label{correlator_l1_s}
\end{eqnarray}
\begin{eqnarray}
&&{\frac{\langle O_{j,l_{1},l_{2}}\chi _{(1^{k})}(Z)\chi
_{(1^{k+j})}(Z)^{\dagger }\rangle }{\sqrt{\langle \chi _{(1^{k+j})}(Z)\chi
_{(1^{k+j})}(Z)^{\dagger }\rangle \langle \chi _{(1^{k})}(Z)\chi
_{(1^{k})}(Z)^{\dagger }\rangle }\left\Vert O_{j,l_{1},l_{2}}\right\Vert }}
\notag \\
&=&\left( {\frac{k}{N}}\right) ^{l_{1}}\left( 1-{\frac{k}{N}}\right) ^{\frac{%
j}{2}+l_{2}}\left( \frac{l+1}{j+2l+1}\right) ^{\frac{1}{2}}.
\label{correlator_l1_a}
\end{eqnarray}

The extremal correlators have been computed \cite{Bissi:2011dc}, see also
related discussion \cite{Caputa:2012yj},\cite{Corley:2001zk}
\begin{eqnarray}
&&{\frac{\langle \mathrm{Tr}(Z^{j})~\chi _{(k)}(Z)\chi _{(k+j)}(Z)^{\dagger
}\rangle }{\sqrt{\langle \chi _{(k+j)}(Z)\chi _{(k+j)}(Z)^{\dagger }\rangle
\langle \chi _{(k)}(Z)\chi _{(k)}(Z)^{\dagger }\rangle }\left\Vert \mathrm{Tr%
}(Z^{j})\right\Vert }}  \notag \\
&=&\left( 1+{\frac{k}{N}}\right) ^{\frac{j}{2}}\frac{1}{\sqrt{j}},
\end{eqnarray}
\begin{eqnarray}
&&{\frac{\langle \mathrm{Tr}(Z^{j})~\chi _{(1^{k})}(Z)\chi
_{(1^{k+j})}(Z)^{\dagger }\rangle }{\sqrt{\langle \chi _{(1^{k+j})}(Z)\chi
_{(1^{k+j})}(Z)^{\dagger }\rangle \langle \chi _{(1^{k})}(Z)\chi
_{(1^{k})}(Z)^{\dagger }\rangle }\left\Vert \mathrm{Tr}(Z^{j})\right\Vert }}
\notag \\
&=&\left( 1-{\frac{k}{N}}\right) ^{\frac{j}{2}}\frac{1}{\sqrt{j}}
\end{eqnarray}%
where we have removed sign factors due to conventions.

In these expressions, the answer for sphere giant and AdS giant are related
by making the replacement $N-k\rightarrow N+k.$ This may also imply that
there are similar limit for sphere giant and AdS giant when ${\frac{k}{N}}$
is fixed but small, which are two alternative ways for the gravitons to blow
up into giant gravitons. The limit we will look at is mainly $k\sim O(N),$
with ${\frac{k}{N}}$ fixed.

%%%%%%%%%%%%%%%%%%%%%%%%%%%%%%%%%%%%%%%%%%%%%%%%%%%%%%%%%%%%%%%%%%%%
%%%%%%%%%%%%%%%%%%%%%%%%%%%%%%%%%%%%%%%%%%%%%%%%%%%%%%%%%%%%%%%%%%%%
%%%%%%%%%%%%%%%%%%%%%%%%%%%%%%%%%%%%%%%%%%%%%%%%%%%%%%%%%%%%%%%%%%%%
%%%%%%%%%%%%%%%%%%%%%%%%%%%%%%%%%%%%%%%%%%%%%%%%%%%%%%%%%%%%%%%%%%%%

\subsection{ Some comparisons}

The correlators with operators for $l=1$ were analysed in detail in \cite%
{Caputa:2012yj}. \ For the $l=2$\ case, there are two types, which gives
different values in the correlator, as explained in the above section. For $%
{\small :}\mathrm{Tr}(Z^{j_{1}}Z^{\dagger }Z^{j-j_{1}+2}Z^{\dagger }){\small %
:}$
\vspace{0.2cm}
\begin{eqnarray}
&&{\frac{\langle {\small :}\mathrm{Tr}(Z^{j_{1}}Z^{\dagger
}Z^{j-j_{1}+2}Z^{\dagger }){\small :}\chi _{(k)}(Z)\chi _{(k+j)}(Z)^{\dagger
}\rangle }{\sqrt{\langle \chi _{(k+j)}(Z)\chi _{(k+j)}(Z)^{\dagger }\rangle
\langle \chi _{(k)}(Z)\chi _{(k)}(Z)^{\dagger }\rangle }\left\Vert {\small :}%
\mathrm{Tr}(Z^{j_{1}}Z^{\dagger }Z^{j-j_{1}+2}Z^{\dagger }){\small :}%
\right\Vert }}  \notag \\
&=&\left( {\frac{k}{N}}\right) ^{2}\left( 1+{\frac{k}{N}}\right) ^{\frac{j}{2%
}}\left( \frac{l+1}{j+2l+1}\right) ^{\frac{1}{2}},
\end{eqnarray}%
with $1\leqslant j_{1}\leqslant j+1.$ On the other hand, for ${\small :}%
\mathrm{Tr}(Z^{j+2}(Z^{\dagger })^{2}){\small :}$%
\vspace{0.2cm}
\begin{eqnarray}
&&{\frac{\langle {\small :}\mathrm{Tr}(Z^{j+2}(Z^{\dagger })^{2}){\small :}%
\chi _{(k)}(Z)\chi _{(k+j)}(Z)^{\dagger }\rangle }{\sqrt{\langle \chi
_{(k+j)}(Z)\chi _{(k+j)}(Z)^{\dagger }\rangle \langle \chi _{(k)}(Z)\chi
_{(k)}(Z)^{\dagger }\rangle }\left\Vert {\small :}\mathrm{Tr}%
(Z^{j+2}(Z^{\dagger })^{2}){\small :}\right\Vert }}  \notag \\
&=&\left( {\frac{k}{N}}\right) \left( 1+{\frac{k}{N}}\right) ^{\frac{j}{2}%
+1}\left( \frac{l+1}{j+2l+1}\right) ^{\frac{1}{2}}.
\end{eqnarray}

\vspace{0.3cm}

We can compare the correlators for two Schur polynomials and one
trace operators with the correlators for three Schur polynomials.
The correlator between three Schur polynomials when one of them is
much smaller than the other two is given by, via
\cite{Corley:2001zk},
\begin{equation}
{\frac{\langle \chi _{(1^{j})}(Z)\chi _{(1^{k})}(Z)\chi
_{(1^{k+j})}(Z)^{\dagger }\rangle }{\sqrt{\langle \chi _{(1^{k+j})}(Z)\chi
_{(1^{k+j})}(Z)^{\dagger }\rangle \langle \chi _{(1^{k})}(Z)\chi
_{(1^{k})}(Z)^{\dagger }\rangle \langle \chi _{(1^{j})}(Z)\chi
_{(1^{j})}(Z)^{\dagger }\rangle }}}=\left( 1-{\frac{k}{N}}\right) ^{\frac{j}{%
2}}
\end{equation}%
where $j\geqslant 1,$ and in the regime $j\ll k~.~$

Denoting $\mathrm{Tr}(Z^{j})=O_{j}~$,
\begin{eqnarray}
&&{\frac{\langle \chi _{(1^{k})}(Z)~O_{j_{1}}O_{j_{2}}~\chi
_{(1^{k+j})}^{\dagger }(Z)\rangle }{\sqrt{\langle \chi _{(1^{k+j})}(Z)\chi
_{(1^{k+j})}(Z)^{\dagger }\rangle \langle \chi _{(1^{k})}(Z)\chi
_{(1^{k})}(Z)^{\dagger }\rangle }\left\Vert O_{j_{1}}\right\Vert \left\Vert
O_{j_{2}}\right\Vert }}  \notag \\
&=&\frac{1}{\sqrt{j_{1}j_{2}}}\left( 1-{\frac{k}{N}}\right) ^{\frac{%
j_{1}+j_{2}}{2}},
\end{eqnarray}%
where $j=j_{1}+j_{2},$
\begin{eqnarray}
&&{\frac{\langle \chi _{(1^{k})}(Z)~O_{j_{1}}O_{j_{2}}...O_{j_{\alpha
}}~\chi _{(1^{k+{}j})}(Z)^{\dagger }\rangle }{\sqrt{\langle \chi
_{(1^{k+j})}(Z)\chi _{(1^{k+j})}(Z)^{\dagger }\rangle \langle \chi
_{(1^{k})}(Z)\chi _{(1^{k})}(Z)^{\dagger }\rangle }\prod\limits_{i}\left%
\Vert O_{j_{i}}\right\Vert }}  \notag \\
&=&\frac{1}{\prod {}_{i=1}^{\alpha }\sqrt{j_{i}}}\left( 1-{\frac{k}{N}}%
\right) ^{\sum {}_{i=1}^{\alpha }\frac{j_{i}}{2}},
\end{eqnarray}%
where $j=\sum {}_{i=1}^{\alpha }j_{i}.$

Correlators between Schur and single trace operators have also
been considered in \cite{Brown:2006zk}, where various transition
processes were explored. See also \cite{Dhar:2005su}. If we
replace one $O_{j_{1}}$ with a light Schur polynomial $\chi
_{(1^{j_{{\small 1}}})}$, then
\begin{eqnarray}
&&{\frac{\langle \chi _{(1^{k})}~\chi _{(1^{j_{1}})}O_{j_{2}}~\chi
_{(1^{k+j})}(Z)^{\dagger }\rangle }{\sqrt{\langle \chi _{(1^{k+j})}(Z)\chi
_{(1^{k+j})}(Z)^{\dagger }\rangle \langle \chi _{(1^{k})}(Z)\chi
_{(1^{k})}(Z)^{\dagger }\rangle \langle \chi _{(1^{j_{1}})}(Z)\chi
_{(1^{j_{1}})}(Z)^{\dagger }\rangle }\left\Vert O_{j_{2}}\right\Vert }}
\notag \\
&=&\frac{1}{\sqrt{j_{2}}}\left( 1-{\frac{j_{1}}{N}}\right) ^{\frac{j_{2}}{2}%
}\left( 1-{\frac{k}{N}}\right) ^{\frac{j_{1}}{2}}
\end{eqnarray}%
for $j_{2}\ll j_{1}$,$~j=j_{1}+j_{2},~$and$~j_{1}\ll k.~$

%%%%%%%%%%%%%%%%%%%%%%%%%%%%%%%%%%%%%%%%%%%%%%%%%%%%%%%%%%%%%%%%%%%%
%%%%%%%%%%%%%%%%%%%%%%%%%%%%%%%%%%%%%%%%%%%%%%%%%%%%%%%%%%%%%%%%%%%%
%%%%%%%%%%%%%%%%%%%%%%%%%%%%%%%%%%%%%%%%%%%%%%%%%%%%%%%%%%%%%%%%%%%%
%%%%%%%%%%%%%%%%%%%%%%%%%%%%%%%%%%%%%%%%%%%%%%%%%%%%%%%%%%%%%%%%%%%%

%%%%%%%%%%%%%%%%%%%%%%%%%%%%%%%%%%%%%%%%%%%%%%%%%%%%%%%%%%%%%%%%%%%%
%%%%%%%%%%%%%%%%%%%%%%%%%%%%%%%%%%%%%%%%%%%%%%%%%%%%%%%%%%%%%%%%%%%%
%%%%%%%%%%%%%%%%%%%%%%%%%%%%%%%%%%%%%%%%%%%%%%%%%%%%%%%%%%%%%%%%%%%%
%%%%%%%%%%%%%%%%%%%%%%%%%%%%%%%%%%%%%%%%%%%%%%%%%%%%%%%%%%%%%%%%%%%%

\section{Correlators with giant gravitons in string theory}

\bigskip

\renewcommand{\theequation}{3.\arabic{equation}} \setcounter{equation}{0}

\subsection{\protect\bigskip Giant graviton computation}

Now let's turn to the discussion from the gravity perspectives. As analysed
in \cite{Bissi:2011dc}, and \cite{Caputa:2012yj}, to compute the correlator
of the giant graviton with a light supergravity field, we consider the
coupling of the supergravity fluctuations to the Euclidean D-brane action
\begin{equation}
S_{D3}=S_{DBI}+S_{WZ}={\frac{N}{2\pi ^{2}}}\int d^{4}\sigma (\sqrt{\mathrm{%
det}(g_{MN}\partial _{a}X^{M}\partial _{b}X^{N})}-iP\left[ C_{4}\right] ).
\end{equation}%
As analysed in \cite{Bissi:2011dc}, we vary the Euclidean D-brane action,
and the DBI part gives the following variation%
\begin{equation}
\delta S_{DBI}=N\cos ^{2}\theta \int dt\,\left( \frac{2}{\Delta +1}Y_{\Delta
}(\partial _{t}^{2}-\Delta ^{2})s^{\Delta }+4\Delta \cos ^{2}\theta
Y_{\Delta }s^{\Delta }\right) ,
\end{equation}%
and the variation of the WZ part is
\begin{equation}
\delta S_{WZ}=N\cos ^{2}\theta \int dt~\left( \left( -4\sin \theta \cos
\theta \partial _{\theta }\right) Y_{\Delta }s^{\Delta }\right) .
\end{equation}%
The total variation of the action is \cite{Bissi:2011dc}
\begin{eqnarray}
&&\delta (S_{DBI}+S_{WZ})  \notag \\
&=&N\cos ^{2}\theta \int dt\,\left( \frac{2}{\Delta +1}Y_{\Delta }(\partial
_{t}^{2}-\Delta ^{2})s^{\Delta }+4(\Delta \cos ^{2}\theta -\sin \theta \cos
\theta \partial _{\theta })Y_{\Delta }s^{\Delta }\right) .  \notag \\
&&
\end{eqnarray}%
Now we can use the expression of the spherical harmonics
\begin{equation}
Y_{\Delta ,\Delta }=\frac{\sin ^{\Delta }\theta }{2^{\Delta /2}}e^{i\Delta
\phi }~  \label{spherical_harmonics_maximal}
\end{equation}%
and replace the field $s^{\Delta }$ with the bulk to boundary propagator
\begin{equation}
s^{\Delta }=\frac{a_{\Delta }z^{\Delta }}{((x-x_{B})^{2}+z^{2})^{\Delta }}%
\rightarrow \frac{a_{\Delta }z^{\Delta }}{x_{B}^{2\Delta }},
\end{equation}%
where $a_{\Delta }=\frac{\Delta +1}{2^{2-\frac{\Delta }{2}}N\sqrt{\Delta }},$
and use the Euclidean geodesic $\phi =\phi (t)=t\rightarrow -it,~z=\frac{R}{%
\cosh t},~x=R\tanh t.$ We also use the notation that $\mathcal{R}%
=R/x_{B}^{2},$ $Y_{\Delta }=\widetilde{Y}_{\Delta }e^{ij\phi }.$

We looked at the integration%
\begin{equation}
\int_{-\infty }^{+\infty }dt\,\frac{e^{jt}}{\cosh ^{\Delta +2n}t}=2^{\Delta
+2n-1}\frac{\Gamma (\frac{1}{2}(\Delta +j)+n)\Gamma (\frac{1}{2}(\Delta
-j)+n)}{\Gamma (\Delta +2n)}.~\ \ \ \
\end{equation}%
The variation is
\begin{eqnarray}
&&\delta S_{DBI}=\frac{a_{\Delta }R^{\Delta }}{x_{B}^{2\Delta }}N\cos
^{2}\theta \left( 4\Delta \cos ^{2}\theta \widetilde{Y}_{\Delta }\cdot
\int_{-\infty }^{+\infty }dt\,\frac{e^{jt}}{\cosh ^{\Delta +2n}t}~\bigg|%
_{n\rightarrow 0}\right.  \notag \\
&&\left. -2\Delta \widetilde{Y}_{\Delta }\cdot \int_{-\infty }^{+\infty }dt\,%
\frac{e^{jt}}{\cosh ^{\Delta +2}t}\right) ,  \label{S_DBI}
\end{eqnarray}

\begin{equation}
\delta S_{WZ}=\frac{a_{\Delta }R^{\Delta }}{x_{B}^{2\Delta }}N\cos
^{2}\theta \left( -4(\sin \theta \cos \theta \partial _{\theta })\widetilde{Y%
}_{\Delta }\cdot \int_{-\infty }^{+\infty }dt\,\frac{e^{jt}}{\cosh ^{\Delta
+2n}t}~\bigg|_{n\rightarrow 0}\right) ,  \label{S_WZ}
\end{equation}%
with the total variation
\begin{eqnarray}
&&\delta S=\frac{a_{\Delta }R^{\Delta }}{x_{B}^{2\Delta }}N\cos ^{2}\theta
\left( 4(\Delta \cos ^{2}\theta -\sin \theta \cos \theta \partial _{\theta })%
\widetilde{Y}_{\Delta }\cdot \int_{-\infty }^{+\infty }dt\,\frac{e^{jt}}{%
\cosh ^{\Delta +2n}t}~\bigg|_{n\rightarrow 0}\right.  \notag \\
&&\left. -2\Delta \widetilde{Y}_{\Delta }\cdot \int_{-\infty }^{+\infty }dt\,%
\frac{e^{jt}}{\cosh ^{\Delta +2}t}\right) .  \label{S_total}
\end{eqnarray}

For extremal correlator, for the maximally charged state,
\begin{equation}
-(\sin \theta \cos \theta \partial _{\theta })\widetilde{Y}_{\Delta
}=-\Delta \cos ^{2}\theta \widetilde{Y}_{\Delta },~~~~\ \ \ \int_{-\infty
}^{+\infty }dt\,\frac{e^{jt}}{\cosh ^{\Delta +2n}t}\bigg|_{n\rightarrow
0}\rightarrow \infty ~.\ \ \ \
\end{equation}%
The first piece of the integral (\ref{S_total}) has a form of $0\cdot \infty
.$

We show a regularization procedure, by first calculating nonextremal
correlator, and then taking a limit $l\rightarrow 0.~$The piece $\Delta \cos
^{2}\theta \widetilde{Y}_{\Delta }$ and $-\sin \theta \cos \theta \partial
_{\theta }\widetilde{Y}_{\Delta }$ are from the DBI and WZ respectively, and
if integrating them separately, one get two divergent terms in (\ref{S_DBI}%
), (\ref{S_WZ}). With the regularization procedure, one can see that their
cancellation is not complete and there is a finite piece when $l\rightarrow 0
$.$~$In this regularization, $4(\Delta \cos ^{2}\theta -\sin \theta \cos
\theta \partial _{\theta })\widetilde{Y}_{\Delta }$ has a factor of $l$,
while $\int_{-\infty }^{+\infty }dt\,\frac{e^{jt}}{\cosh ^{\Delta +2n}t}$
has a factor $\Gamma (l).$

For the spherical harmonics,
\begin{equation}
Y_{\Delta ,\Delta -2l}=\frac{\Gamma (j+l+1)\sqrt{(j+l+1)(l+1)}2^{-j/2}}{%
\Gamma (l+2)\Gamma (j+1)\sqrt{j+2l+1}\ 2^{l}}\sin ^{j}\theta e^{i(j\phi
)}~_{2}F_{1}(-l,j+l+2,j+1;\sin ^{2}\theta ),  \label{sp_harmonic}
\end{equation}%
with $l=0,1,2...$

Let's define%
\begin{equation}
F_{\Delta ,j}=\sin ^{j}\theta ~_{2}F_{1}(-l,j+l+2,j+1;\sin ^{2}\theta ),
\end{equation}%
in which $_{2}F_{1}(-l,j+l+2,j+1;\sin ^{2}\theta )|_{l=0}=1.$

\begin{eqnarray}
\Delta \cos ^{2}\theta F_{\Delta ,j} &=&j\cos ^{2}\theta F_{\Delta
,j}+2l\cos ^{2}\theta F_{\Delta ,j},  \notag \\
-\sin \theta \cos \theta \partial _{\theta }F_{\Delta ,j} &=&-j\cos
^{2}\theta F_{\Delta ,j}+2l\cos ^{2}\theta \sin ^{j+2}\theta ~\frac{j+l+2}{%
j+1}~_{2}F_{1}(1-l,j+l+3,j+2;\sin ^{2}\theta ).  \notag \\
&&
\end{eqnarray}%
Those two first terms in (\ref{S_DBI}),(\ref{S_WZ}) when integrated are
divergent when $l=0$. On the other hand the integral has an overall $\Gamma
(l)~$factor:
\begin{equation}
\int_{-\infty }^{+\infty }dt\,\frac{e^{jt}}{\cosh ^{j+2l}t}=2^{j+2l-1}\frac{%
\Gamma (j+l)}{\Gamma (j+2l)}\Gamma (l)
\end{equation}%
\newline
which is divergent at $l=0.$

On the other hand,
\begin{eqnarray}
&&\cos ^{2}\theta (F_{\Delta ,j}+\sin ^{j+2}\theta ~\frac{j+l+2}{j+1}%
~_{2}F_{1}(1-l,j+l+3,j+2;\sin ^{2}\theta ))|_{l\rightarrow 0}  \notag \\
&=&\frac{\sin ^{j}\theta }{\cos ^{2}\theta }\frac{1}{j+1}(1+j\cos ^{2}\theta
).  \label{jcos^2}
\end{eqnarray}

Now if we add the two pieces together, and using $l\Gamma (l)|_{l=0}=1$ when
setting $l=0$, we find%
\begin{eqnarray}
&&\frac{a_{\Delta }R^{\Delta }}{x_{B}^{2\Delta }}N\cos ^{2}\theta ~4(\Delta
\cos ^{2}\theta -\sin \theta \cos \theta \partial _{\theta })\widetilde{Y}%
_{\Delta }\int_{-\infty }^{+\infty }dt\,\frac{e^{jt}}{\cosh ^{j+2l}t}~\bigg|%
_{l\rightarrow 0}  \notag \\
&=&(2\mathcal{R})^{\Delta }\frac{1}{\sqrt{j}}\left( 1+{\frac{k}{N}j}\right)
\left( 1-{\frac{k}{N}}\right) ^{\frac{j}{2}}  \label{piece_r}
\end{eqnarray}%
where the factor $1+{\frac{k}{N}j}$ is from the factor $1+j\cos ^{2}\theta $
in (\ref{jcos^2}). The two divergent terms therefore do not cancel
completely and yield a finite piece.

The other piece in the integral is
\begin{equation}
\frac{a_{\Delta }R^{\Delta }}{x_{B}^{2\Delta }}N\cos ^{2}\theta (-2\Delta
\widetilde{Y}_{\Delta }\cdot \int_{-\infty }^{+\infty }dt\,\frac{e^{jt}}{%
\cosh ^{\Delta +2}t})=-(2\mathcal{R})^{\Delta }{\frac{k}{N}}\sqrt{j}\left( 1-%
{\frac{k}{N}}\right) ^{\frac{j}{2}}.  \label{piece_n}
\end{equation}

Now adding all the pieces together from (\ref{piece_r}),(\ref{piece_n}), we
find~that the coefficient is
\begin{equation}
-(2\mathcal{R})^{\Delta }(-1)^{j-1}\frac{1}{\sqrt{j}}\left( 1-{\frac{k}{N}}%
\right) ^{\frac{j}{2}}
\end{equation}%
where we reinstated the $(-1)^{j}$ factor from section \ref%
{sec_correlator_01} due to conventions. This agrees with gauge theory
computation,\

\begin{equation}
(-1)^{j-1}\frac{1}{\sqrt{j}}\left( 1-{\frac{k}{N}}\right) ^{\frac{j}{2}}.
\end{equation}

The computation above indicates that the previous mismatch discussed in \cite%
{Bissi:2011dc}\ is mainly due to the subtlety of extremal correlators. There
were two possible reasons in understanding the mismatch: One is that \cite%
{Bissi:2011dc} interpreted this disagreement as the Schur polynomials'
inability to interpolate between giant gravitons and pointlike gravitons;
while another is that \cite{Caputa:2012yj} thought that the problem was
related to subtleties in extremal correlators. Here, the above computation
shows that the previous discrepancy of the computation between gauge theory
and string theory is related to the subtlety of the extremal correlators.

%%%%%%%%%%%%%%%%%%%%%%%%%%%%%%%%%%%%%%%%%%%%%%%%%%%%%%%%%%%%%%%%%%%%
%%%%%%%%%%%%%%%%%%%%%%%%%%%%%%%%%%%%%%%%%%%%%%%%%%%%%%%%%%%%%%%%%%%%
%%%%%%%%%%%%%%%%%%%%%%%%%%%%%%%%%%%%%%%%%%%%%%%%%%%%%%%%%%%%%%%%%%%%
%%%%%%%%%%%%%%%%%%%%%%%%%%%%%%%%%%%%%%%%%%%%%%%%%%%%%%%%%%%%%%%%%%%%

%%%%%%%%%%%%%%%%%%%%%%%%%%%%%%%%%%%%%%%%%%%%%%%%%%%%%%%%%%%%%%%%%%%%
%%%%%%%%%%%%%%%%%%%%%%%%%%%%%%%%%%%%%%%%%%%%%%%%%%%%%%%%%%%%%%%%%%%%
%%%%%%%%%%%%%%%%%%%%%%%%%%%%%%%%%%%%%%%%%%%%%%%%%%%%%%%%%%%%%%%%%%%%
%%%%%%%%%%%%%%%%%%%%%%%%%%%%%%%%%%%%%%%%%%%%%%%%%%%%%%%%%%%%%%%%%%%%

\subsection{Sphere giant and comparisons}

\bigskip \label{sec_giant_02}

Let's first consider the sphere giant. Let's consider the general expression
of $Y^{\Delta ,\Delta -2l}$ in (\ref{sp_harmonic}). For example, for $l=1,$
\begin{equation}
Y_{\Delta ,j}=\frac{\sqrt{2(j+2)}(j+1)2^{-j/2}}{4\sqrt{j+3}}\sin ^{j}\theta
e^{i(j\phi )}~(1-\frac{j+3}{j+1}\sin ^{2}\theta ).
\end{equation}%
These situations have been considered in detail in \cite{Caputa:2012yj}, and
perfect agreement were found there.

We now calculate the answer for other $l.~$We use the equality
\begin{eqnarray}
&&(\cos ^{2}\theta -\frac{j+l}{j+2l+1})_{2}F_{1}(-l,j+l+2,j+1;\sin
^{2}\theta )  \notag \\
&&+\cos ^{2}\theta \sin ^{2}\theta ~\frac{j+l+2}{j+1}%
~_{2}F_{1}(1-l,j+l+3,j+2;\sin ^{2}\theta )  \notag \\
&=&\frac{l+1}{j+2l+1}~_{2}F_{1}(1-l,j+l+1,j+1;\sin ^{2}\theta )=\frac{l+1}{%
j+2l+1}\cos ^{{\small -2}}\theta ~_{2}F_{1}(-l,j+l,j+1;\sin ^{2}\theta ).
\notag \\
&&
\end{eqnarray}

We then have the expression
\begin{equation}
\delta S=(2\mathcal{R})^{\Delta }\frac{\Gamma (j+l+1)\Gamma (j+l)\sqrt{%
(j+l+1)(l+1)}}{\Gamma (j+2l)\Gamma (j+1)\sqrt{(j+2l)(j+2l+1)}}\sin
^{j}\theta ~_{2}F_{1}(-l,j+l,j+1;\sin ^{2}\theta ).
\end{equation}

The coefficient is

\begin{eqnarray}
&&\frac{\langle O_{(j+2l,j)}\chi _{(1^{k})}\chi _{(1^{k+j})}^{\dagger
}\rangle }{\sqrt{\langle \chi _{(1^{k+j})}\chi _{(1^{k+j})}^{\dagger
}\rangle \langle \chi _{(1^{k})}\chi _{(1^{k})}^{\dagger }\rangle }%
\left\Vert O_{(j+2l,j)}\right\Vert }  \notag \\
&=&\frac{\Gamma (j+l+1)\Gamma (j+l)\sqrt{(j+l+1)(l+1)}}{\Gamma (j+2l)\Gamma
(j+1)\sqrt{(j+2l)(j+2l+1)}}\left( 1-{\frac{k}{N}}\right) ^{\frac{j}{2}%
}~_{2}F_{1}(-l,j+l,j+1;1-{\frac{k}{N}})  \notag \\
&&
\end{eqnarray}%
where we have removed the $(-1)^{j-1}$ factors due to conventions.

For $l=0,\ _{2}F_{1}(0,j,j+1;\sin ^{2}\theta )=1,$%
\begin{equation}
\frac{\langle O_{(j,j)}\chi _{(1^{k})}\chi _{(1^{k+j})}^{\dagger }\rangle }{%
\sqrt{\langle \chi _{(1^{k+j})}\chi _{(1^{k+j})}^{\dagger }\rangle \langle
\chi _{(1^{k})}\chi _{(1^{k})}^{\dagger }\rangle }\left\Vert
O_{(j,j)}\right\Vert }=\frac{1}{\sqrt{j}}\left( 1-{\frac{k}{N}}\right) ^{%
\frac{j}{2}}
\end{equation}%
where we have removed a $(-1)^{j-1}$ factor due to conventions. This agrees
with the gauge theory computation.

For $l=1,\ _{2}F_{1}(-1,j+1,j+1;\sin ^{2}\theta )=\cos ^{2}\theta ,$%
\begin{equation}
\frac{\langle O_{(j+2,j)}\chi _{(1^{k})}\chi _{(1^{k+j})}^{\dagger }\rangle
}{\sqrt{\langle \chi _{(1^{k+j})}\chi _{(1^{k+j})}^{\dagger }\rangle \langle
\chi _{(1^{k})}\chi _{(1^{k})}^{\dagger }\rangle }\left\Vert
O_{(j+2,j)}\right\Vert }={\frac{k}{N}}\left( 1-{\frac{k}{N}}\right) ^{\frac{j%
}{2}}\left( \frac{2}{j+3}\right) ^{\frac{1}{2}}.
\end{equation}%
Those were considered in \cite{Caputa:2012yj}.

For sphere giant, when $\frac{k}{N}\rightarrow 1,$ the correlator is very
small, with the exception of $j=0~$case; on the gravity side, this is
because of that the giant has to make a change of the R-charge, while $\frac{%
k}{N}\rightarrow 1$ makes it difficult to do so, so it is suppressed due to
R-charge conservation of the process. For non-extremal correlators, with
neutral operators, we see that it is not small even when $1-\frac{k}{N}$
becomes small, for the special case $j=0$;~on the gravity side, this is
because the giant does not need to make a change of the R-charge when it is
maximal.

For $l\geqslant 2,$ in the limit $l^{2}\ll j,~_{2}F_{1}(-l,j+l,j+1;\sin
^{2}\theta )=\cos ^{2l}\theta (1+o(l^{2}/j))$ for $l\geqslant 2,$
\begin{equation}
\frac{\langle O_{(j+2l,j)}\chi _{(1^{k})}\chi _{(1^{k+j})}^{\dagger }\rangle
}{\sqrt{\langle \chi _{(1^{k+j})}\chi _{(1^{k+j})}^{\dagger }\rangle \langle
\chi _{(1^{k})}\chi _{(1^{k})}^{\dagger }\rangle }\left\Vert
O_{(j+2l,j)}\right\Vert }=\left( {\frac{k}{N}}\right) ^{l}\left( 1-{\frac{k}{%
N}}\right) ^{\frac{j}{2}}\left( \frac{l+1}{j+2l+1}\right) ^{\frac{1}{2}%
}(1+o(l^{2}/j)).
\end{equation}%
There is simplification in the limit $l^{2}/j\ll 1.$ In this limit the
leading order functional dependence on ${\frac{k}{N}}$ is the same as in (%
\ref{correlator_rep_1k_non_a}) for the situation without adjacent $%
Z^{\dagger }$'s. This means that in this limit, the main light states are
those with non-adjacent $Z^{\dagger }$'s. This limit is consistent with the
gauge theory computation, and it means that the number of $ZZ^{\dagger }$'s
is much fewer than the $Z$'s in the trace.

In the limit $l^{2}\ll j,$ most of the operators that appears in (\ref%
{operator_Z+_ge}) are the operators with non-adjacent $Z^{\dagger }.$
Therefore this limit extract the term with the largest power of $\frac{k}{N}$%
, which is $\left( \frac{k}{N}\right) ^{l}$, in the polynomial given by the
hypergeometric function, giving the $\left( \frac{k}{N}\right) ^{l}\left( 1-{%
\frac{k}{N}}\right) ^{\frac{j}{2}}$. Other terms with smaller powers of $%
\frac{k}{N}$ are from the operators with adjacent $Z^{\dagger }$'s. In the
expansion of the hypergeometric function, the terms of order $%
o(l^{2}/j)^{\alpha },$ has the form $\left( \frac{k}{N}\right) ^{l-\alpha
}\left( 1-{\frac{k}{N}}\right) ^{\frac{j}{2}+\alpha }.$ We think that those
terms with $o(l^{2}/j)^{\alpha }$ are from the operators with non-adjacent $%
Z^{\dagger },$ in which the number of zero $j_{i}$'s in the expression (\ref%
{operator_Z+_ge}) is $\alpha .~$If there is $\alpha $ zero $j_{i}$'s$,$ the
number of such terms goes as $\binom{j+l-\alpha -1}{j}\binom{l}{\alpha },$
while the number of terms with no adjacent $Z^{\dagger }$'s goes as $\binom{%
j+l-1}{j}$. The ratio of their numbers approach the coefficients in the
expansion of the hypergeometric function
\begin{eqnarray}
&&\left( 1-{\frac{k}{N}}\right) ^{\frac{j}{2}}~_{2}F_{1}(-l,j+l,j+1;1-{\frac{%
k}{N}})  \notag \\
&=&\sum_{\alpha =0}^{l-1}\frac{(-l)_{\alpha }(1-l)_{\alpha }}{\alpha
!(j+1)_{\alpha }}\left( \frac{k}{N}\right) ^{l-\alpha }\left( 1-{\frac{k}{N}}%
\right) ^{\frac{j}{2}+\alpha }\ ,
\end{eqnarray}%
where in the $l^{2}\ll j$ limit, $\ r_{\alpha }=\frac{(-l)_{\alpha
}(1-l)_{\alpha }}{\alpha !(j+1)_{\alpha }}\sim \binom{j+l-\alpha -1}{j}%
\binom{l}{\alpha }\binom{j+l-1}{j}^{-1}\sim o(l^{2}/j)^{\alpha }.$

We see that the $l=0,l=1$ cases all agree. The operators are%
\begin{equation}
O_{(j,j)}=\mathrm{Tr}(Z^{j})
\end{equation}%
\begin{equation}
O_{(j+2,j)}={\small :}\mathrm{Tr}(Z^{j+1}Z^{\dagger }){\small :.}
\end{equation}

For $l\geqslant 2,$ the correlators have different values, according to how
many non-adjacent $Z^{\dagger }$'s are, see discussions of (\ref%
{correlator_l1_a}). The result in the small $l^{2}/j$ limit is the
limit where most of the terms are those with non-adjacent
$Z^{\dagger }$.

%The result in the small $l^{2}/j$ limit is the limit where the
%operator with non-adjacent $Z^{\dagger }$ has the most number.

For $l=2$,
\begin{equation}
\frac{\langle O_{(j+4,j)}\chi _{(1^{k})}\chi _{(1^{k+j})}^{\dagger }\rangle
}{\sqrt{\langle \chi _{(1^{k+j})}\chi _{(1^{k+j})}^{\dagger }\rangle \langle
\chi _{(1^{k})}\chi _{(1^{k})}^{\dagger }\rangle }\left\Vert
O_{(j+4,j)}\right\Vert }=\frac{\sqrt{3(j+3)}}{\sqrt{(j+4)(j+5)}}\left( 1-{%
\frac{k}{N}}\right) ^{\frac{j}{2}}{\frac{k}{N}}\left( {\frac{k}{N}-}\frac{{2}%
}{j+3}\right) .
\end{equation}%
This implies the combination from the gauge theory computation%
\begin{equation}
O_{(j+4,j)}=\frac{1}{2}\sum_{j_{1}=1}^{j+1}{\small :}\mathrm{Tr}%
(Z^{j_{1}}Z^{\dagger }Z^{j+2-j_{1}}Z^{\dagger }){\small :~}+~{\small :}%
\mathrm{Tr}(Z^{j+2}Z^{\dagger }Z^{\dagger }){\small :.}  \label{l=2_op}
\end{equation}%
The factor of $\frac{1}{2}$ is because that the operator with $j_{1}$ and $%
j_{1}\rightarrow j+2-j_{1}$ is the same by cyclicity. Now we see that there
are two different type of terms, one with $Z^{\dagger }$'s non-adjacent, and
one with $Z^{\dagger }$'s adjacent. As in section \ref{sec_correlator_01},
the $\frac{k}{N}$ dependence of these two types in the correlators are
different, one with $\left( \frac{k}{N}\right) ^{2}\left( 1-{\frac{k}{N}}%
\right) ^{\frac{j}{2}}~$and another with $-\frac{k}{N}\left( 1-\frac{k}{N}%
\right) \left( 1-{\frac{k}{N}}\right) ^{\frac{j}{2}}.$ Note that there are $%
j+1$ terms of the first type with no adjacent $Z^{\dagger },~$in the
summation in the first part of (\ref{l=2_op}), so the normalized answer is
\begin{eqnarray}
&&\frac{\langle O_{(j+4,j)}\chi _{(1^{k})}\chi _{(1^{k+j})}^{\dagger
}\rangle }{\sqrt{\langle \chi _{(1^{k+j})}\chi _{(1^{k+j})}^{\dagger
}\rangle \langle \chi _{(1^{k})}\chi _{(1^{k})}^{\dagger }\rangle }%
\left\Vert O_{(j+4,j)}\right\Vert }  \notag \\
&=&\frac{\sqrt{3}(j+1)}{\sqrt{(j+3)(j+4)(j+5)}}\left( 1-{\frac{k}{N}}\right)
^{\frac{j}{2}}\left( \left( \frac{k}{N}\right) ^{2}{-}\frac{{2}}{j+1}\frac{k%
}{N}\left( 1-\frac{k}{N}\right) \right) .  \label{correlator_l=2_s_gg}
\end{eqnarray}%
The two types of terms in (\ref{l=2_op}) gives the two terms in (\ref%
{correlator_l=2_s_gg}), and are distinguishable.

We can study an opposite limit. Let's study the $j=0$ limit.
\begin{equation}
\frac{\langle O_{(2l,0)}\chi _{(1^{k})}\chi _{(1^{k})}^{\dagger }\rangle }{%
\sqrt{\langle \chi _{(1^{k})}\chi _{(1^{k})}^{\dagger }\rangle \langle \chi
_{(1^{k})}\chi _{(1^{k})}^{\dagger }\rangle }\left\Vert
O_{(2l,0)}\right\Vert }=\frac{\Gamma (l)\Gamma (l+2)}{\Gamma (2l)\sqrt{%
2l(2l+1)}}~_{2}F_{1}(-l,l,1;1-\frac{k}{N}).
\end{equation}

For $l=2$, it is
\begin{equation}
\frac{\langle O_{(4,0)}\chi _{(1^{k})}\chi _{(1^{k})}^{\dagger }\rangle }{%
\sqrt{\langle \chi _{(1^{k})}\chi _{(1^{k})}^{\dagger }\rangle \langle \chi
_{(1^{k})}\chi _{(1^{k})}^{\dagger }\rangle }\left\Vert O_{(4,0)}\right\Vert
}=\frac{3}{2\sqrt{5}}{\frac{k}{N}}\left( {\frac{k}{N}-}\frac{{2}}{3}\right) .
\end{equation}

This can be evaluated using the method in \cite{Skenderis:2007yb}, and \cite%
{Caputa:2012yj}, via the integration, and for the sphere giant
computation, for $j=0,$ it is
\begin{equation}
\langle O^{(4,0)}\rangle =\int_{0}^{\infty }\,{\frac{N~e^{-Nr^{2}}(Nr^{2})^{%
\frac{-2k-2+2N}{2}}}{\sqrt{5}\sqrt{(N-k-1)!(N-k-1)!}}}\,(3r^{4}-4r^{2}+1)%
\,rdr=\frac{3}{2\sqrt{5}}\left( \left( {\frac{k}{N}}\right) ^{2}{-}\frac{{2}%
}{3}{\frac{k}{N}}\right)  \label{O_4,0_1k}
\end{equation}%
where we rescaled $\pi a^{2}$~factors. This is for the $(1^{k})$
representation. %The neutral operators match very well.
The analysis of $%
\langle O^{(4,2)}\rangle ,\langle O^{(3,1)}\rangle ,\langle O^{(2,0)}\rangle
$ has been performed in \cite{Caputa:2012yj}, in agreement with both the
field theory result and the anatomy of \cite{Skenderis:2007yb}, and they are
both consistent with \cite{Lin:2004nb}.

%%%%%%%%%%%%%%%%%%%%%%%%%%%%%%%%%%%%%%%%%%%%%%%%%%%%%%%%%%%%%%%%%%%%
%%%%%%%%%%%%%%%%%%%%%%%%%%%%%%%%%%%%%%%%%%%%%%%%%%%%%%%%%%%%%%%%%%%%
%%%%%%%%%%%%%%%%%%%%%%%%%%%%%%%%%%%%%%%%%%%%%%%%%%%%%%%%%%%%%%%%%%%%
%%%%%%%%%%%%%%%%%%%%%%%%%%%%%%%%%%%%%%%%%%%%%%%%%%%%%%%%%%%%%%%%%%%%

%%%%%%%%%%%%%%%%%%%%%%%%%%%%%%%%%%%%%%%%%%%%%%%%%%%%%%%%%%%%%%%%%%%%
%%%%%%%%%%%%%%%%%%%%%%%%%%%%%%%%%%%%%%%%%%%%%%%%%%%%%%%%%%%%%%%%%%%%
%%%%%%%%%%%%%%%%%%%%%%%%%%%%%%%%%%%%%%%%%%%%%%%%%%%%%%%%%%%%%%%%%%%%
%%%%%%%%%%%%%%%%%%%%%%%%%%%%%%%%%%%%%%%%%%%%%%%%%%%%%%%%%%%%%%%%%%%%

\subsection{AdS giant and comparisons}

In the case of AdS giant, in this particular situation $\theta
=\pi /2.$ The variations have been
analysed in detail \cite{Bissi:2011dc}. The possibility of arbitrary $%
Y_{\Delta ,j}$ are considered in \cite{Caputa:2012yj}. The expression with
the $t$ integral and summation of the series is \cite{Caputa:2012yj},\cite%
{Bissi:2011dc}
\begin{equation}
\delta S=-\frac{2^{\frac{\Delta }{2}-1}\sqrt{\Delta }}{\Gamma (\Delta +1)}%
\frac{\mathcal{R}^{\Delta }}{\cosh ^{\Delta }\rho }\sum_{n=0}^{\infty }\frac{%
1}{2^{2n}}\frac{\Gamma (\Delta +2n+2)}{\Gamma (n+2)\Gamma (n+1)}\tanh
^{2n+2}\rho \int_{-\infty }^{\infty }dt\frac{Y_{\Delta ,j}}{\cosh ^{\Delta
+2+2n}t}.  \label{summation}
\end{equation}%
We have included also $l\geqslant 2$ cases, as well as general $j$ for $l=1$
cases.

Performing the summation in (\ref{summation}), the expression is
\begin{eqnarray}
\delta S &=&-(2\mathcal{R})^{\Delta }\frac{\Gamma (j+l+1)\sqrt{(j+l+1)(l+1)}%
}{\Gamma (j+2l)\sqrt{(j+2l)(j+2l+1)}}(1-x)^{\frac{j}{2}+l}d_{x}^{l}\left(
\frac{x^{l+1}}{l+1}~_{2}F_{1}(l+1,j+l+1,l+2;x)+\frac{1}{j}\right)  \notag \\
&&
\end{eqnarray}%
and
\begin{eqnarray}
&&\frac{\langle O_{(j+2l,j)}\chi _{(k)}\chi _{(k+j)}^{\dagger }\rangle }{%
\sqrt{\langle \chi _{(k+j)}\chi _{(k+j)}^{\dagger }\rangle \langle \chi
_{(k)}\chi _{(k)}^{\dagger }\rangle }\left\Vert O_{(j+2l,j)}\right\Vert }
\notag \\
&=&\frac{\Gamma (j+l+1)\sqrt{(j+l+1)(l+1)}}{\Gamma (j+2l)\sqrt{(j+2l)(j+2l+1)%
}}(1-x)^{\frac{j}{2}+l}d_{x}^{l}\left( \frac{x^{l+1}}{l+1}%
~_{2}F_{1}(l+1,j+l+1,l+2;x)+\frac{1}{j}\right) ,  \notag \\
&&
\end{eqnarray}%
where $x=\frac{k}{N+k}=\tanh ^{2}\rho .$

\begin{eqnarray}
&&\lim_{l\rightarrow 0}\frac{\Gamma (j+l+1)\sqrt{(j+l+1)(l+1)}}{\Gamma (j+2l)%
\sqrt{(j+2l)(j+2l+1)}}(1-x)^{\frac{j}{2}+l}d_{x}^{l}\left( \frac{x^{l+1}}{l+1%
}~_{2}F_{1}(l+1,j+l+1,l+2;x)+\frac{1}{j}\right)  \notag \\
~ &=&\ \ \frac{1}{\sqrt{j}}(1-x)^{-\frac{j}{2}}.~\ \ ~
\end{eqnarray}

By first computing the answer for general $l$ and then taking the limit $%
l\rightarrow 0$, the answer calculated in this way agrees with the gauge
theory computation.

For $l=0,~x~_{2}F_{1}(1,j+1,2;x)+\frac{1}{j}=\frac{1}{j}(1-x)^{-j},\delta
S=-(2\mathcal{R})^{\Delta }\frac{1}{\sqrt{j}}(1-x)^{-\frac{j}{2}},$
\begin{equation}
\frac{\langle O_{(j,j)}\chi _{(k)}\chi _{(k+j)}^{\dagger }\rangle }{\sqrt{%
\langle \chi _{(k+j)}\chi _{(k+j)}^{\dagger }\rangle \langle \chi _{(k)}\chi
_{(k)}^{\dagger }\rangle }\left\Vert O_{(j,j)}\right\Vert }=\left( 1+{\frac{k%
}{N}}\right) ^{\frac{j}{2}}\frac{1}{\sqrt{j}}.
\end{equation}

For $l=1,$
\begin{equation}
\frac{\langle O_{(j+2,j)}\chi _{(k)}\chi _{(k+j)}^{\dagger }\rangle }{\sqrt{%
\langle \chi _{(k+j)}\chi _{(k+j)}^{\dagger }\rangle \langle \chi _{(k)}\chi
_{(k)}^{\dagger }\rangle }\left\Vert O_{(j+2,j)}\right\Vert }={\frac{k}{N}}%
\left( 1+{\frac{k}{N}}\right) ^{\frac{j}{2}}\left( \frac{2}{j+3}\right) ^{%
\frac{1}{2}},
\end{equation}%
which were analysed in detail in \cite{Caputa:2012yj}.

For $l\geqslant 2,$ in the limit $l^{2}\ll j,$
\begin{eqnarray}
&&(1-x)^{\frac{j}{2}+l}d_{x}^{l}\left( \frac{x^{l+1}}{l+1}%
~_{2}F_{1}(l+1,j+l+1,l+2;x)+\frac{1}{j}\right)   \notag \\
&=&\frac{\Gamma (j+2l)}{\Gamma (j+l+1)}\frac{x^{l}}{(1-x)^{\frac{j}{2}+l}}%
(1+o(l^{2}/j)).
\end{eqnarray}%
So in the limit~$l^{2}\ll j,$%
\begin{equation}
\frac{\langle O_{(j+2l,j)}\chi _{(k)}\chi _{(k+j)}^{\dagger }\rangle }{\sqrt{%
\langle \chi _{(k+j)}\chi _{(k+j)}^{\dagger }\rangle \langle \chi _{(k)}\chi
_{(k)}^{\dagger }\rangle }\left\Vert O_{(j+2l,j)}\right\Vert }=\left( {\frac{%
k}{N}}\right) ^{l}\left( 1+{\frac{k}{N}}\right) ^{\frac{j}{2}}\left( \frac{%
l+1}{j+2l+1}\right) ^{\frac{1}{2}}(1+o(l^{2}/j)).
\label{correlator_ads_gg_l>2}
\end{equation}%
Note that for $l\geqslant 2,~o(l^{2}/j)$ is no smaller than $o(l/j).~$The
leading order expression (\ref{correlator_ads_gg_l>2}) is the approximation
when $l^{2}\ll j.$ This is in agreement with (\ref{correlator_rep_k_non_a}),
and see related discussion in section \ref{sec_giant_02}.

For $l=2$,
\begin{equation}
\frac{\langle O_{(j+4,j)}\chi _{({k})}\chi _{({k+j})}^{\dagger
}\rangle }{\sqrt{\langle \chi _{({k+j})}\chi _{({k+j})}^{\dagger
}\rangle \langle \chi _{({k})}\chi _{({k})}^{\dagger }\rangle
}\left\Vert
O_{(j+4,j)}\right\Vert }=\frac{\sqrt{3(j+3)}}{\sqrt{(j+4)(j+5)}}\left( 1+{%
\frac{k}{N}}\right) ^{\frac{j}{2}}{\frac{k}{N}}\left( {\frac{k}{N}+}\frac{{2}%
}{j+3}\right) .
\end{equation}%
This again agrees with the discussion in section \ref{sec_giant_02}.

We can also study the $j=0$ limit. For $l=2$, it is
\begin{equation}
\frac{\langle O_{(4,0)}\chi _{(k)}\chi _{(k)}^{\dagger }\rangle }{\sqrt{%
\langle \chi _{(k)}\chi _{(k)}^{\dagger }\rangle \langle \chi _{(k)}\chi
_{(k)}^{\dagger }\rangle }\left\Vert O_{(4,0)}\right\Vert }=\frac{3}{2\sqrt{5%
}}{\frac{k}{N}}\left( {\frac{k}{N}+}\frac{{2}}{3}\right) .
\end{equation}%
Again using the analysis \cite{Skenderis:2007yb}, and \cite{Caputa:2012yj},
\begin{equation}
\langle O^{(4,0)}\rangle =\int_{0}^{\infty }\,{\frac{N~e^{-Nr^{2}}(Nr^{2})^{%
\frac{2k-2+2N}{2}}}{\sqrt{5}\sqrt{(N+k-1)!(N+k-1)!}}}\,(3r^{4}-4r^{2}+1)%
\,rdr=\frac{3}{2\sqrt{5}}\left( \left( {\frac{k}{N}}\right) ^{2}{+}\frac{{2}%
}{3}{\frac{k}{N}}\right)  \label{O_4,0,k}
\end{equation}%
which is for the $(k)$ representation, and with $j=0$. Situations
are quite similar to section \ref{sec_giant_02}.

%%%%%%%%%%%%%%%%%%%%%%%%%%%%%%%%%%%%%%%%%%%%%%%%%%%%%%%%%%%%%%%%%%%%
%%%%%%%%%%%%%%%%%%%%%%%%%%%%%%%%%%%%%%%%%%%%%%%%%%%%%%%%%%%%%%%%%%%%
%%%%%%%%%%%%%%%%%%%%%%%%%%%%%%%%%%%%%%%%%%%%%%%%%%%%%%%%%%%%%%%%%%%%
%%%%%%%%%%%%%%%%%%%%%%%%%%%%%%%%%%%%%%%%%%%%%%%%%%%%%%%%%%%%%%%%%%%%

%%%%%%%%%%%%%%%%%%%%%%%%%%%%%%%%%%%%%%%%%%%%%%%%%%%%%%%%%%%%%%%%%%%%
%%%%%%%%%%%%%%%%%%%%%%%%%%%%%%%%%%%%%%%%%%%%%%%%%%%%%%%%%%%%%%%%%%%%
%%%%%%%%%%%%%%%%%%%%%%%%%%%%%%%%%%%%%%%%%%%%%%%%%%%%%%%%%%%%%%%%%%%%
%%%%%%%%%%%%%%%%%%%%%%%%%%%%%%%%%%%%%%%%%%%%%%%%%%%%%%%%%%%%%%%%%%%%

\section{Discussions}

\renewcommand{\theequation}{9.\arabic{equation}} \setcounter{equation}{0}

\label{discussion}

We studied the correlators between two heavy operators of the Schur
polynomial type and one light operator, from the gauge theory point of view.
Both non-extremal correlators and extremal correlators are studied.

In the situation of the extremal correlator, we showed a regularization
procedure in which the string theory computation matched the gauge theory
computation exactly. We first calculate the expression for small $l$ and
then extends it to $l=0$. In the $l\rightarrow 0$ limit, both the DBI and WZ
variation have a property that they contain divergent terms, and their
cancellation is not completely zero. The regularization procedure gives a
finite answer, which agrees with the gauge theory computation exactly.

We also studied non-extremal correlators with $l=1$, but with general
R-charge $j$, and they also agree with the gauge theory computation for
general R-charge.

We analysed the non-extremal correlators with general $l$, including $%
l\geqslant 2.~$In the $l^{2}\ll j$ limit, the gravity answers are simplified
there. We also showed that in an opposite limit, $j=0$ limit, with $l=2$
example$,$ the correlator also agrees, with both the gauge theory
computation and another method using anatomy of \cite{Skenderis:2007yb}. In
particular, $\ $nontrivial $k/N$ dependence is nicely captured by the
gravity computations.

From the gravity point of view, there are certain similarities
with the computation for the case of string worldsheet and
supergravity field, see for example
\cite{Zarembo:2010rr}-\cite{Bak:2011yy}. There might be
connections between these approaches.

\section*{Acknowledgments}

\vspace{1pt}

This work is supported in part by Xunta de Galicia (Conselleria de Educacion
and grants PGIDIT10PXIB 206075PR and INCITE09 206121PR), by the Spanish
Consolider-Ingenio 2010 Programme CPAN (CSD2007-00042), by the Juan de la
Cierva of MICINN, by the ME, MICINN and Feder (grant FPA2011-22594), by the
UNIFY Programme on Frontiers in Theoretical Physics, and the Seventh
Framework Programme (grant n%
%TCIMACRO{\U{b0}}%
%BeginExpansion
${{}^\circ}$%
%EndExpansion
269217). We also would like to thank Caltech, and Isaac Newton Institute for
Mathematical Sciences, Cambridge, for hospitalities.

%%%%%%%%%%%%%%%%%%%%%%%%%%%%%%%%%%%%%%%%%%%%%%%%%%%%%%%%%%%%%%%%%%%%
%%%%%%%%%%%%%%%%%%%%%%%%%%%%%%%%%%%%%%%%%%%%%%%%%%%%%%%%%%%%%%%%%%%%
%%%%%%%%%%%%%%%%%%%%%%%%%%%%%%%%%%%%%%%%%%%%%%%%%%%%%%%%%%%%%%%%%%%%
%%%%%%%%%%%%%%%%%%%%%%%%%%%%%%%%%%%%%%%%%%%%%%%%%%%%%%%%%%%%%%%%%%%%
%%%%%%%%%%%%%%%%%%%%%%%%%%%%%%%%%%%%%%%%%%%%%%%%%%%%%%%%%%%%%%%%%%%%

\appendix

\renewcommand{\theequation}{A.\arabic{equation}} \setcounter{equation}{0}

\vspace{1pt}

\vspace{1pt}

\vspace{1pt}%%%%%%%%%%%%%%%%%%%%%%%%%%%%%%%%%%%%%%%%%%%%%%%%%%%%%%%%

\vspace{1pt}

\vspace{1pt}%\cite{Maldacena:1997re}

%%%%%%%%%%%%%%%%%%%%%%%%%%%%%%%%%%%%%%%%%%%%%%%%%%%%%%%%%%%%%%%%%%%%
%%%%%%%%%%%%%%%%%%%%%%%%%%%%%%%%%%%%%%%%%%%%%%%%%%%%%%%%%%%%%%%%%%%%
%%%%%%%%%%%%%%%%%%%%%%%%%%%%%%%%%%%%%%%%%%%%%%%%%%%%%%%%%%%%%%%%%%%%
%%%%%%%%%%%%%%%%%%%%%%%%%%%%%%%%%%%%%%%%%%%%%%%%%%%%%%%%%%%%%%%%%%%%

%%%%%%%%%%%%%%%%%%%%%%%%%%%%%%%%%%%%%%%%%%%%%%%%%%%%%%%%%%%%%%%%%%%%
%%%%%%%%%%%%%%%%%%%%%%%%%%%%%%%%%%%%%%%%%%%%%%%%%%%%%%%%%%%%%%%%%%%%
%%%%%%%%%%%%%%%%%%%%%%%%%%%%%%%%%%%%%%%%%%%%%%%%%%%%%%%%%%%%%%%%%%%%
%%%%%%%%%%%%%%%%%%%%%%%%%%%%%%%%%%%%%%%%%%%%%%%%%%%%%%%%%%%%%%%%%%%%
\vspace{1pt}

\vspace{1pt}

\end{document}